\documentclass[aps,prl,twocolumn]{revtex4}
\begin{document}



\title{A Singlet-pairing superconductor is always also a super-spin-current-conductor} %
\author{Chia-Ren Hu}
\email{crhu@tamu.edu}
\affiliation{Department of Physics, Texas A\&M University, College Station, TX
77843-4242, USA}
%
%
\date{November 12, 2007}
\begin{abstract}
It is shown that, in a conductor carrying a moderate spin current, singlet pairing can 
still take place almost as effectively as without a spin current. The system will still 
be fully gapped for all spin-up and -down single-particle excitations, with no depaired
electrons. All thermodynamic properties will be practically the same as without the
spin current. All universality relations and laws of corresponding states found in the
theory of Bardeen, Cooper, and Schrieffer should remain valid. Thus a singlet-pairing 
superconductor can always carry a persistent and dissipation-less spin-current. 
A heuristic argument to support this conclusion is also given, as well as two possible 
experimental tests of this prediction.
\end{abstract}
\maketitle

The Bardeen, Cooper, Schrieffer (BCS) theory of superconductivity\cite{BCS} is often naively
introduced as Bose-Einstein condensation (BEC) of bound pairs of electrons in the
spin-singlet channel, which is represented by an antisymmetric spin wave-function of one 
spin-up electron and one spin-down electron.  Breaking such a bound pair would cost twice 
of an energy gap, so it seems impossible for the two electrons to go separate ways without 
first obtaining this energy. If this naive picture were correct, it would be inconceivable 
for the condensate of a singlet-pairing superconductor (SPSC) to be able to carry a dc 
spin current, which is defined as the spin-up and -down electrons flowing in opposite 
directions (relative to some spin quantization axis), with no net charge flow. 
Fortunately the BCS state is much more subtle than that. Thus, as an important difference 
between a singlet-pairing BCS state and a BEC state of tightly bound, 
spin-singlet fermion pairs, we show below that in a conductor carrying a moderate 
spin current, spin-singlet Cooper pairing\cite{Cooper} of electrons can still take 
place almost as effectively as without the spin current. An energy gap still 
opens up on both the spin-up- and -down Fermi surfaces, which are now centered at 
$\pm{\bf q}/2$ momentum, and are no longer in overlap.~\cite{note-swave} 
Transition temperature and energy gap are only extremely weakly reduced from those in the 
absence of a spin current, and all universality relations and laws of corresponding states 
found in the BCS theory should remain valid.  Since the imposed spin current is found to
co-exist with spin-singlet pairing at absolute zero temperature, when the system is fully 
gapped and no excitations can possibly exist, we conclude that a spin current can flow 
persistently and without dissipation in a SPSC. The spin current should thus be called a 
super-spin-current (SSpC), and the system, a super-spin-current-conductor
(SSpCC). Two experimental tests of this prediction are proposed below.

First, we offer a heuristic picture: Consider $n$ dancing couples arranged in a big circle 
with each couple circling a small circle at diametrically opposite positions like a bound 
state of two electrons. If each couple is also advancing along the big circle, 
it would be the analog of a super(-charge-)current (SChC). If all dancers now agree that, 
in every five minutes, say, changing partners {\it coherently} should take place, 
with each man stepping forward and each woman stepping backward, along the big circle, until 
new dancing partners can be formed, one then finds that a new kind of flow is occurring in the 
floor, with all men advancing in the counterclockwise direction, and all women in the clockwise 
direction. This is an analog of a SpC, with a man (woman) the analog 
of a spin-up (-down) electron. If the intra-pair distance were smaller than half of the the 
inter-pair distance, the pairs would have to be broken before new pairs can be formed. 
In the BCS state of a SC, however, the intra-pair distance ---
which is of the order of the coherence length~\cite{Cooper} $\xi$ --- is much larger than the 
inter-pair distance, so that Cooper-pair orbits are in strong overlap, allowing this 
partner-changing process to take place without the need to break the pairs, and thus 
a SSpC can flow in the system. This partner-changing process would be drastically 
suppressed in a BEC state of tightly bound pairs, where the intra-pair distance 
is much smaller than the inter-pair distance, although quantum mechanical wave-functions do 
not cut off abruptly, so the probability that this process can happen is never zero. Thus we 
surmise that only an extremely weak SSpC can exist in such a BEC state, to the extend of 
being negligible.

For a simple theory, we begin with the second-quantized BCS Hamiltonian:
\begin{eqnarray}
{\cal K}_0 &\equiv& {\cal H}_0 - \mu N = \sum_{{\bf k}, s}\xi_{\bf k}
    \hat c^{\dagger}_{{\bf k}, s}\hat c_{{\bf k}, s} - \nonumber \\
    & &\hspace{-0.5in}(\lambda/2\Omega)
    \sum_{{\bf k}_1, {\bf k}_2,{\bf q}, s_1, s_2}\hat c^{\dagger}_{{\bf k}_1 + 
    {\bf q}, s_1}\hat c^{\dagger}_{{\bf k}_2 - {\bf q}, s_2}
    \hat c_{{\bf k}_2, s_2}\hat c_{{\bf k}_1, s_1}\,,
\end{eqnarray}
where $\Omega$ the total volume of the system, $\lambda > 0$ an effective coupling 
constant, and $\xi_{\bf k}\equiv \hbar^2{\bf k}^2/2m^* - \mu$, with $m^*$ the 
electron effective mass, and $\mu$ the chemical potential.  Adding to it the 
Zeeman Hamiltonian:
\begin{equation}
{\cal H}_1 = -h\sum_{{\bf k}, s, s'}\hat c^{\dagger}_{{\bf k}, s}\sigma^z_{ss'}
             \hat c_{{\bf k}, s'}\,,
\end{equation}
(where ${\sigma}^z$ is the third Pauli matrix, and $h$ is the electron magnetic
moment times an external magnetic field in the $z$ direction, and can be viewed 
as a Lagrange multiplier,) would favor a spin imbalance along $z$ in the spin 
space.  Similarly, to favor a ChC, one can introduce a vector 
Lagrange multiplier ${\bf v}_{ch}$, whose direction defines the direction of the 
charge flow, and add to ${\cal H}_0$:
\begin{equation}
{\cal H}_2 = -{\bf v}_{ch}\cdot\sum_{{\bf k}, s}\hbar{\bf k}\hat c^{\dagger}_{{\bf k}, s}
   \hat c_{{\bf k}, s} + (1/2)m{\bf v}_{ch}^2\sum_{{\bf k}, s}\hat c^{\dagger}_{{\bf k}, s}
   \hat c_{{\bf k}, s}\,,
\end{equation}
where the second term is a correction to $\mu$. Thus to favor a SpC in the system
(with the same spin quantization direction), we introduce a vector 
Lagrange multiplier ${\bf v}_{sp}$, and add to ${\cal H}_0$:
\begin{equation}
{\cal H}_3 = -{\bf v}_{sp}\cdot\sum_{{\bf k}, s, s'}\hbar{\bf k}
     \hat c^{\dagger}_{{\bf k}, s}\sigma^z_{ss'}\hat c_{{\bf k}, s'} + 
     (1/2)m{\bf v}_{sp}^2\sum_{{\bf k}, s}\hat c^{\dagger}_{{\bf k}, s}
     \hat c_{{\bf k}, s}\,.
\end{equation} 

For discussing a superconductor carrying no SChC, BCS neglected all interaction matrix 
elements that involves electron-pairs of non-zero momentum. 
Then a mean-field approximation changes ${\cal K}_0$ to: 
\begin{eqnarray}
{\cal K}_{0,{\mathrm MF}} &=& \sum_{{\bf k}, s}\xi_{\bf k}\hat c^{\dagger}_{{\bf k}, s}
    \hat c_{{\bf k}, s} - \Delta^*\sum_{{\bf k}}\hat c_{{-\bf k}, \downarrow}
  \hat c_{{\bf k}, \uparrow}\nonumber\\ 
  &-& \Delta\sum_{\bf k}\hat c^{\dagger}_{{\bf k}, \uparrow}
  \hat c^{\dagger}_{-{\bf k}, \downarrow} + (\Omega/\lambda)|\Delta|^2\,,
\end{eqnarray}
\begin{equation}
\Delta\equiv (\lambda/\Omega)\sum_{\bf k}<<\hat c_{{-\bf k}, \downarrow}
              \hat c_{{\bf k}, \uparrow}>>_T\,,
\label{Delta}
\end{equation}
with $<<\cdots>>_T$ denoting grand-canonical ensemble average at temperature $T$.
The Hamiltonian ${\cal K}_{3, {\mathrm MF}}\equiv {\cal K}_{0, {\mathrm MF}} + 
{\cal H}_3$ can be diagonalized in the same way as ${\cal K}_{0, MF}$:  
by using a Bogoliubov-Valatin transformation:
\begin{eqnarray}
\hat c_{{\bf k}, \uparrow}&=&u^*_{{\bf q},{\bf k}}\hat\gamma_{{\bf q},{\bf k},\uparrow}
     +v_{{\bf q},{\bf k}}\hat\gamma^{\dagger}_{{\bf q},-{\bf k},\downarrow} \\
\hat c^{\dagger}_{-{\bf k}, \downarrow}&=& -v^*_{{\bf q},{\bf k}}
     \hat\gamma_{{\bf q},{\bf k},\uparrow}
     +u_{{\bf q},{\bf k}}\hat\gamma^{\dagger}_{{\bf q},-{\bf k},\downarrow}\,,
\end{eqnarray}
where ${\bf q}\equiv 2m^*{\bf v}_{sp}/\hbar$.  The result is:
\begin{eqnarray}
{\cal K}_{3, {\mathrm MF}}&=&\sum_{\bf k}E_{{\bf q},{\bf k}}
(\hat\gamma^{\dagger}_{{\bf q},{\bf k},\uparrow}\hat\gamma_{{\bf q},{\bf k},\uparrow}
+\hat\gamma^{\dagger}_{{\bf q},-{\bf k},\downarrow}\hat\gamma_{{\bf q},-{\bf k},\downarrow})
\nonumber\\
&+&\sum_{\bf k}(\xi_{{\bf q},{\bf k}}-E_{{\bf q},{\bf k}})+(\Omega/\lambda)|\Delta_{\bf q}|^2\,,
\end{eqnarray}
\begin{eqnarray}
u_{{\bf q},{\bf k}}&=& \left[ \frac{1}{2}(1+\frac{\xi_{{\bf q},{\bf k}}}
     {E_{{\bf q},{\bf k}}})\right]^{\frac{1}{2}}\\
v_{{\bf q},{\bf k}}&=& \frac{\Delta_{\bf q}}{|\Delta_{\bf q}|}
     \left[ \frac{1}{2}(1-\frac{\xi_{{\bf q},{\bf k}}} 
     {E_{{\bf q},{\bf k}}})\right]^{\frac{1}{2}}\,,
\end{eqnarray}
\begin{equation}
\xi_{{\bf q},{\bf k}}\equiv \xi_{({\bf k} - {\bf q}/2)}\,,
\end{equation}
\begin{equation}
E_{{\bf q},{\bf k}}=\sqrt{\xi_{{\bf q},{\bf k}}^2+|\Delta_{\bf q}|^2}\,,
\end{equation}
where $\Delta_{\bf q}$ is defined similar to $\Delta$ of Eq.~\ref{Delta}.  That 
$E_{{\bf q},{\bf k}}$ is positive definite for all ${\bf q}$ shows that a SpC 
does not induce pair breaking in the system.

Many physical quantities can now be evaluated as in the original BCS theory:  
The difference between the superconducting and normal ground-state energies at $T = 0$ is:
\begin{eqnarray}
\Delta E_3|_{T=0}&\equiv& <<{\cal K}^{(S)}_{3,MF}>>_{T=0} - <<{\cal K}^{(N)}_{3,MF}>>_{T=0}
\nonumber\\ 
&=& -(1/2){\cal N}(0)\Omega|\Delta_{{\bf q},0}|^2\,,
\end{eqnarray}
where ${\cal N}(0)$ is the density of states per spin at the ${\bf q} = 0$ Fermi 
energy, and the usual weak-coupling approximation has been assumed; $\Delta_{{\bf q},0}\equiv 
\Delta_{\bf q}|_{T=0}$. Let the cut-off energies be still $\pm\hbar \omega_c$ for energy 
integration around the (${\bf q}=0$) Fermi energy, the zero-temperature gap equation
\begin{eqnarray}
1 &=& (1/2){\cal N}(0)\lambda\int_{-1}^{1}\frac{d\mu}{2}\left[\ln\frac{2(\hbar\omega_c - 
     \hbar v_F q\mu)}{\Delta_{{\bf q},0}} + \right. \nonumber\\
  & &\left.\hspace{0.5in}\ln\frac{2(\hbar\omega_c + \hbar v_F q\mu)}{\Delta_{{\bf q},0}}\right] 
\end{eqnarray}
gives:
\begin{equation}
\Delta_{{\bf q},0}\simeq\Delta_{{\bf 0},0}
     \exp{\left [-\frac{1}{6}\left(\frac{\hbar v_F q/2}{\hbar\omega_c}\right)^2\right]}\,,
\label{Deltaq0}
\end{equation}
with $\Delta_{{\bf 0},0}$ the $T=0$ gap in the BCS theory. The gap equation at $T\ne 0$:
\begin{equation}
1 = \frac{\lambda}{\Omega}\sum_{\bf k}\frac{1}{2E_{{\bf q},{\bf k}}}
    \tanh\left( \frac{E_{{\bf q},{\bf k}}}{2k_B T}\right)
\label{GapEq}    
\end{equation}
implies	
\begin{equation}
T_{c,{\bf q}} = T_{c,{\bf 0}}
    \exp{\left[-\frac{1}{6}\left(\frac{\hbar v_F q/2}{\hbar\omega_c}\right)^2\right]}\,,
\label{Tcq}
\end{equation}
where $T_{c,{\bf 0}}$ is the transition temperature in the BCS theory. Thus 
the universality relation of the BCS theory is obtained:
\begin{equation}
\frac{2\Delta_{{\bf q},0}}{k_B T_{c,{\bf q}}} = 2\pi e^C = 3.528\,,
\end{equation}
with $C = 0.577$ the Euler constant. From Eq.~\ref{GapEq}, one can derive the same 
BCS universality relation between $|\Delta_{\bf q}/\Delta_{{\bf q},0}|$ and $T/T_{c,{\bf q}}$. 
The fractional specific-heat jump at $T_{c,{\bf q}}$, 
or $\Delta C/C_n|_{T_{c,{\bf q}}}$, is also found to be that 
of the BCS theory. Other thermodynamic quantities have not been investigated, 
but we expect all universality relations and laws of corresponding states in the 
BCS theory to remain valid.  Evaluating the ensemble average of the SpC
density:
\begin{eqnarray}
\langle\langle{\bf j}_{sp}\rangle\rangle_T&\equiv& \nonumber\\
& &\hspace{-0.8 in}(1/\Omega)
\langle\langle\sum_{\bf k}(\hbar {\bf k}/m^*)
   (\hat c^{\dagger}_{{\bf k},\uparrow}\hat c_{{\bf k},\uparrow} + 
   \hat c^{\dagger}_{-{\bf k},\downarrow}\hat c_{-{\bf k},\downarrow})\rangle\rangle_T\nonumber\\
& &\hspace{-0.8 in} = \frac{1}{\Omega}\sum_{\bf k}\frac{\hbar{\bf k}}{m^*} 
   \left(1-\frac{\xi_{{\bf q},{\bf k}}}{E_{{\bf q},{\bf k}}}
   \tanh\frac{E_{{\bf q},{\bf k}}}{2k_BT}\right)\,,
\label{jsp}
\end{eqnarray}
which, in the weak coupling approximation, is always equal to its normal-state value 
$n{\bf v}_{sp}$ at all $T<T_{c,{\bf q}}$ (and above), where $n$ is the 
electron density, and ${\bf v}_{sp}$ is the spin-current velocity.  
This finite SpC must be all SSpC at $T=0$, because at 
$T=0$ there is no normal fluid component and the system is fully gapped, and it must 
be all normal SpC at $T=T_{c,{\bf q}}$, because we have a continuous phase 
transition to the normal state at this temperature.  At any $T$ between 0 and 
$T_{c, {\bf q}}$ it must be partly super and partly normal, but we have not yet split 
apart the two components. The important point established here is that a 
SSpC exists and that a SPSC can carry a SSpC.  If the corresponding ChC problem 
${\cal K}_{0, MF} + {\cal H}_2$ is solved by the same approach, and the ensemble-average 
of the ChC density is calculated, one would obtain   
\begin{equation}
\langle\langle{\bf j}\rangle\rangle_T\, = \frac{2e\hbar}{m}\frac{1}{\Omega}
  \sum_{\bf k}\,{\bf k}\,f(-\frac{\hbar{\bf k}}{m^*}\cdot\frac{{\bf q}}{2} + E_{\bf k})\,,
\end{equation}
(${\bf q}\equiv 2m^*{\bf v}_{ch}/\hbar$) which vanishes at $T = 0$ and is equal to the 
normal-state value $ne{\bf v}_{ch}$ at $T = T_c$. This ChC is all normal at any $T$: 
The Hamiltonian ${\cal K}_{0, MF} + {\cal H}_2$ is 
just ${\cal K}_{0, MF}$ in the frame moving with the velocity ${\bf v}_{ch}$. 
In that frame the normal fluid component is stationary, whereas the pair condensate is 
moving with velocity $-{\bf v}_{ch}$. Thus $\langle\langle{\bf j}\rangle\rangle_T$ in 
that frame is all SChC. Translating back to the lab frame, we obtain only normal current,  
since the pair condensate is now stationary. Such a trick of going to a moving frame 
can not be played to the SpC problem. Our mean-field solution for 
${\cal K}_{0, MF} + {\cal H}_3$ is also not a mean-field solution of 
${\cal K}_{0, MF}$ alone, which might bother some readers, as it seems that the 
``fictitious field'' ${\bf v}_{sp}$ cannot be realized in the laboratory.~\cite{note-vsp}  
Our reply is: We think that a superconducting sample subject to an external SpC 
(in the $x$ direction, say) can be modeled as the Hamiltonian ${\cal K}_0$ together 
with the boundary condition 
$
\hat\psi_{\sigma}({\bf x} + L\hat e_x)= e^{-i(q/2)L\sigma}\hat\psi_{\sigma}({\bf x})
\label{BC}
$ 
for the field operator $\hat\psi_{\sigma}({\bf x})$. If one then makes a gauge 
transformation 
$
\hat\psi_{\sigma}({\bf x})=e^{-i(q/2)x\sigma}\hat\psi'_{\sigma}({\bf x})\,,
$
one would obtain ${\cal K}_{0} + {\cal H}_3$ with ${\bf v}_{sp}$ along $x$ 
and the usual periodic boundary conditions, which we have solved. 
This gauge transformation changes the eigen-wave-functions, 
but not the eigen-energies, nor $\Delta_{\bf q}$. 

We expect $v_{sp}$ to be comparable in magnitude to $v_{ch}$, and is of the order of 
1 mm/s or less. Thus  
$\hbar v_F (q/2)/\hbar\omega_c = m^*v_Fv_{sp}/\hbar\omega_c$ appearing in Eqs.~\ref{Deltaq0} 
and \ref{Tcq} is of the order of $10^{-7}$ or smaller. Thus 
$\Delta_{\bf q,0}/\Delta_{0,0}$ and $T_{c, {\bf q}}/T_{c,0}$ are exceedingly close to unity. 
We conclude that all thermodynamic properties of a superconductor are practically not 
affected by the presence of a laboratory-realizable SpC in the system. 

Increasing the magnitude of a SChC velocity can lead to two critical values: 
At the first, Landau critical velocity, depairing begins, and at the second one 
the order parameter is suppressed to zero~\cite{depairing} (in three dimensions 
--- In $<3$ dimensions the two critical velocities merge into one~\cite{zhang}). 
No depairing can be induced by a SSpC velocity as long as weak-coupling approximation 
is valid. So no critical SSpC velocity can be defined this way.  Still, 
a critical SSpC velocity likely exists for $\hbar v_F q > \hbar \omega_c$, but 
its value would be so large that it is practically irrelevant. The Landau 
critical velocity is already quite large, but for SChC, there exist lower 
critical velocities due to the creation of phase-slip centers in one dimension 
and super-current vortices (or flux lines) in higher dimensions. The decay 
mechanisms for SSpC carried by a SPSC remain to be investigated,\cite{note} 
but it appears that a SPSC can at least carry comparable SSpC as SChC 
(after artificially introducing a factor of $|e|$ to the definition in 
Eq.~\ref{jsp} so they can be compared).  This is potentially useful to 
the development of SpC circuits and for spin injection, as it appears 
that SpC can not flow in most normal conductors for macroscopic distances.  

If the length $L$ is that of the circumference of a ring sample, then  
$(q/2)L = 2n\pi$ for an integer $n$, so that $\hat\psi_{\sigma}({\bf x})$ 
can satisfy the usual periodic boundary conditions.  The SSpC is then quantized 
as a SChC in such a situation. Thus line singularities in SSpC can also exist,
as vortices (flux lines) in SChC, even as phase gradient of the pair wave 
function is no longer a relevant concept to SSpC.  

The ultimate confirmation of this prediction must be via experimentation. 
We thus propose the following two experimental tests of our prediction:      

In the first proposed test, one should simply connect a SPSC to a battery 
except that at least one section of the connection wiring should be made 
of a half-metallic conductor (HMC)~\cite{halfmetal}, allowing, say, only 
spin-up current. If the SPSC can carry a dc SSpC (referred to as ``scenario A''), 
then it can also carry a dc super-spin-up-current, since an equal mixing of a 
SpC and a ChC is just a pure spin-up current. No voltage drop will occur across 
the SC, so the whole voltage drop will exist in the leads, driving the same 
amount of pure spin-up current in the leads. On the other hand, if the SPSC 
can not carry a dc SSpC (referred to as ``scenario B''), then spin-up current 
cannot flow through the SC, and will initially generate spin-up charge accumulations 
at the interfaces between the leads and the SC. Spin-neutral SChC will then be 
induced in the SC, until the spin-up charge accumulations at the said interfaces 
are all converted to pure spin-density accumulations. The SChC will then stop. 
The ``SpC voltage''~\cite{note2} established in the leads will then cancel out 
the effect of the battery voltage on the spin-up charges in the lead, (but they 
will add up on spin-down charges which, however, can not flow in the leads,)
so that no net electro-chemical-potential gradient will be acting on the 
spin-up electrons in the leads, and no net voltage will be acting across the 
SC. The final state is a zero current state. 
The difference between these two scenarios can be easily differentiated by a 
current meter in the circuit, or by the magnetic field generated by the spin-up 
current in the scenario A only.  If the HMCs 
are only nearly perfect, then some small ChC will always flow in the circuit, 
\textsc{}which becomes SChC in the SC, but the 
actual total current flowing in the circuit will be much larger in scenario A 
than in scenario B.  In scenario A the proper ratio of spin-up and -down 
currents will flow in the circuit, as if the SC does not exist in the circuit, 
whereas in scenario B, the final state is such that only a small amount 
of spin-neutral ChC will flow in the circuit, which becomes pure 
SChC in the SC.
Spin-density accumulations will exist at the said interfaces, to reduce the 
electro-chemical-potential gradient of the spin-up electrons, and enhance that 
of the spin-down electrons, by so much so that spin-up and -down current densities 
in the circuit can be equal. This is a self consistency problem which will be 
investigated in the future.      
  
In the second proposed test, a SC strip is shaped into a ring with a 
narrow gap, which is filled with a HMC allowing only spin-down 
current, with good contacts with both ends of the SC strip 
to form a closed circuit of low resistance. A battery with two HMC leads connected 
to the two ends of the SC strip are used to send a pure spin-up current  
through the SC strip. In scenario B no ChC or SpC will flow 
through the SC. In scenario A the battery will succeed in sending a spin-up 
current through the SC strip, but to reduce the magnetic energy associated 
with the magnetic field induced in the loop, we think that a spin-down current will  
be induced in the ring, so that in the SC there is only a SSpC 
but no SChC. To distinguish between these two scenarios one 
can use either a current meter to detect the spin-up current in 
scenario A only, or a thermometer placed in contact with the HMC 
filling the gap of the ring in order to detect the heat generated by the spin-down 
current in it in scenario A only. In either scenarios there is little magnetic field 
in the ring to be detected. Imperfect HMCs can be similarly analyzed.

Hirsch has pointed out~\cite{Hirsch-spincurrent} that if the BCS wave function is 
altered to break parity but not the time-reversal symmetry, it can then carry 
a dissipation-less SSpC. Thus the ground-state wave-function 
obtained here as a SpC-carrying wave-function is not new, but as the mean-field  
ground-state wave-function of a meaningful Hamiltonian associated with a SpC in the 
system it is new.  In addition, the main purpose of Ref.~\cite{Hirsch-spincurrent} is 
to discuss when a symmetry-breaking SCing (i.e., pairing) 
condensate will carry also a spontaneous SpC. Thus it has to find special model 
Hamiltonians to achieve this goal.  On the other hand, the main purpose of the present 
work is to discuss what happens if one attempts to send a pure SpC through an ordinary, 
SPSC. We conclude that {\it all} SPSCs 
can let a SpC flow through without dissipation, and the SCing 
properties of such a system is little affected by any moderate SpC flowing 
in it. Furthermore, a quantized, persistent, dissipation-less SSpC
can exist in any ring sample of any ordinary SPSC.  No special 
Hamiltonian is needed to achieve either goals. We have also proposed 
two currently feasible experimental tests of this prediction.  Spin-orbit 
interaction has not yet been considered in this work. This interaction led
Hirsch~\cite{Hirsch-Meissner} to propose a spin Meissner effect as a SC 
response to ionic electric fields.

The concept of SSpCC (or SpC SC) has also been introduced in a very different 
context in spin-orbit-split hole-bands of semiconductors such as Si or GaAs, 
without invoking pairing at all.~\cite{semicond-SSCC-1,semicond-SSCC-2} An 
{\it off-diagonal} transport coefficient was considered there, which is naturally 
dissipation-less. But when diagonal transport coefficients are considered at 
the same time, the SpC may no longer be dissipation-less. The SSpC introduced in 
the present work is associated with a diagonal transport coefficient --- SpC 
conductivity: If a dc SpC can exist persistently and without dissipation, this 
diagonal transport coefficient must diverge in the zero-frequency limit, just 
like the dc (ChC-)conductivity in a SC. Thus the concept of SSpCC introduced 
here is genuine. Its potential for practical applications should be enormous.      

The author wishes to thank J. H. Ross, Jr., and Yong Chen for useful comments. 
The author acknowledges some summer support from the Texas Center for 
Superconductivity at the University of Houston in 2007.



\begin{thebibliography}{99}
\bibitem{note-swave} An $s$-wave pairing is assumed here. Extention to non-s-wave singlet 
pairing is straightforward. We have also not yet investigated the possibility of a small 
admixture of higher waves (such as d-wave) to an originally s-wave pairing state before 
a spin-current is introduced.
\bibitem{note} Triplet up-up and down-down pairings can easily carry a spin current, 
but phase-slip-centers or vortices can also exist independently in each spin species, 
leading to similar magnitudes for the critical SpC and ChC velocities.
\bibitem{note2} The ``SpC electric field'' is the sum of a chemical potential gradient that is 
opposite for spin-up and -down electrons, and, possibly, a Zeeman-field gradient.
\end{thebibliography}
\end{document}